\documentclass{article}
\usepackage{spconf,amsmath,graphicx,hyperref}
\usepackage{amssymb}
\usepackage{booktabs}
\usepackage{multirow}
\usepackage{adjustbox}
\usepackage{float}
\usepackage{makecell}

\title{SPEAR-Gen: Generation-Aware Pre-training for Unified Speech Representations}
\name{\vspace{-0.5em}Xiaoyu Yang\thanks{$^*$Work done while at Meta Superintelligence Labs as an intern}$^{\ddagger *}$, Arthur Hinsvark$^{\dagger}$, Antonios Alexos$^{\dagger}$, Osama Hanna$^{\dagger}$, Philip C. Woodland$^{\ddagger}$, Yiting Lu$^{\dagger}$}
\address{$^{\ddagger}$Department of Engineering, University of Cambridge, Cambridge, UK\\$^{\dagger}$Meta Superintelligence Labs, Menlo Park, USA\vspace{-0.5em}}
\begin{document}
%
\maketitle

\begin{abstract}

Speech understanding and generation place different demands on speech representations, and existing models are typically optimised towards one capability or the other.
To reduce this gap, we introduce SPEAR-Gen, a speech representation model that learns a single representation for both capabilities. 
Task-aligned feature aggregation consolidates complementary linguistic and paralinguistic information across a frozen encoder into discrete targets for masked prediction, while a coarse-to-fine objective combines log-Mel reconstruction with residual flow matching to preserve spectral structure and fine-grained acoustic variation.
Experiments on SUPERB and speech resynthesis show that SPEAR-Gen maintains strong understanding performance while substantially improving resynthesis quality and speaker preservation. These results demonstrate that a single speech representation can effectively support both understanding and generation.


\end{abstract}
\vspace{-0.2em}
\section{Introduction}
\label{sec:intro}
\vspace{-0.2em}

Speech conveys linguistic content together with rich paralinguistic and acoustic cues. Understanding-oriented learning often prioritises linguistic content~\cite{hubert,wavlm,spear}, whereas generation additionally requires these acoustic cues to be faithfully preserved~\cite{soundstream,encodec}. Existing representation models therefore tend to specialise in one direction or maintain separate feature spaces.
A single representation that makes these complementary forms of information jointly accessible is highly desirable, providing a common interface for diverse speech applications and future speech language models~\cite{speechgpt,moshi,QwenOmni}.

However, learning such a unified representation remains challenging because existing training objectives impose different representational biases. 
Masked token prediction (MTP) trains an encoder output to predict discrete targets obtained by clustering features from a particular layer in a pre-trained model~\cite{hubert,wavlm,spear}. Such targets can inherit layer-specific biases, encouraging the output representation to specialise in phonetic or linguistic content while leaving speaker and other paralinguistic cues more accessible from earlier layers~\cite{SslModelEvalTaslp}. 
On the other hand, reconstruction objectives preserve rich acoustic information but show weak understanding performance~\cite{mockingjay,tera,encodec}. 
Although recent studies have sought to bridge these gaps~\cite{olive,uniwav,WavCube}, they either rely on separate representations for understanding and generation or retain a trade-off between both.

\begin{figure*}[!h]
    \centering
    \includegraphics[width=\linewidth]{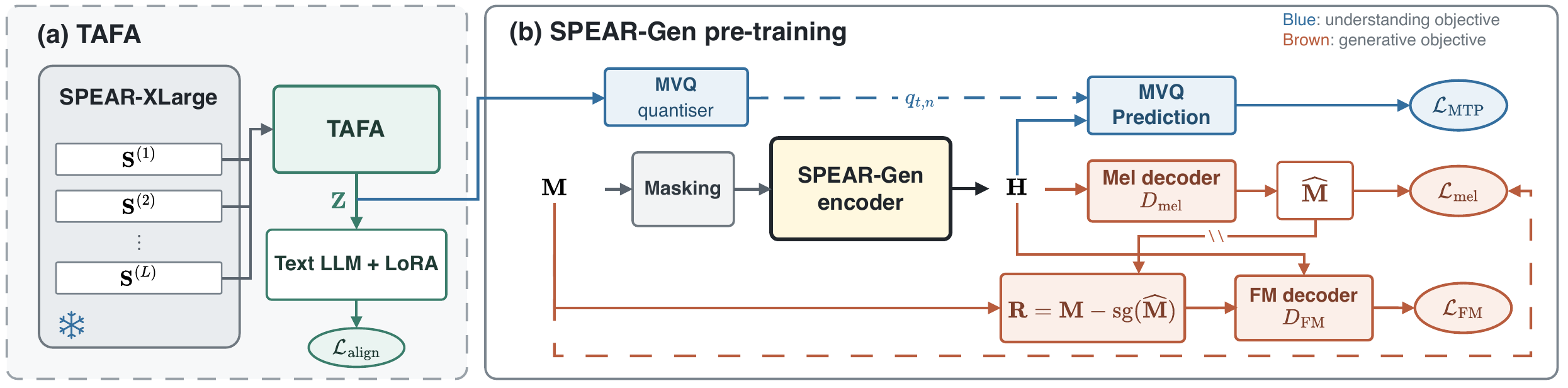}
    \vspace{-2.2em}
    \caption{Overview of SPEAR-Gen. (a) TAFA fuses layer-wise features from SPEAR-XLarge under multi-task supervision, and its output is quantised into MVQ targets. (b) SPEAR-Gen jointly learns masked MVQ prediction and two-stage coarse-to-fine generative modelling through log-Mel reconstruction and residual FM. Only the SPEAR-Gen encoder is retained.}
    \label{fig:spear_gen}
    \vspace{-0.8em}
\end{figure*}

To address this gap, we introduce \textbf{SPEAR-Gen} (Figure~\ref{fig:spear_gen}), a speech representation model that learns a single shared representation for understanding and generation. 
First, task-aligned feature aggregation (TAFA) uses task supervision to consolidate linguistic and paralinguistic cues across a frozen encoder into a task-aligned representation that is quantised into MVQ targets, guiding MTP to expose these cues jointly to the final layer representation.
Second, the coarse-to-fine generative modelling combines mel reconstruction with residual flow matching (FM). Mel reconstruction encourages the representation to retain dominant spectro-temporal structure, while residual FM provides complementary supervision for the remaining nuanced acoustic variation.

Experiments on the SUPERB benchmark~\cite{superb} and speech resynthesis demonstrate that SPEAR-Gen achieves a strong balance between speech understanding and generation within a single-layer representation.
Compared with understanding-focused counterparts, adding generation-aware training substantially improves speech resynthesis while preserving linguistic performance and strengthening paralinguistic information. Compared with WavCube~\cite{WavCube}, SPEAR-Gen achieves competitive resynthesis quality while showing stronger understanding performance.
These results suggest that a single representation can support both speech understanding and generation, achieving a favourable balance.



\vspace{-1em}
\section{Related Work}
\label{sec:related}
\vspace{-0.4em}

SPEAR~\cite{spear} learns understanding-oriented speech representations through masked prediction of MVQ~\cite{MVQ} tokens. SPEAR-Gen builds on SPEAR, but constructs its MVQ targets through fused representations across multiple layers and introduces generative modelling to support both understanding and generation.
Recent studies have explored joint learning for speech understanding and generation. OLIVE~\cite{olive} combines latent prediction with waveform reconstruction, but uses different feature levels for understanding and synthesis. UniWav~\cite{uniwav} incorporates flow-matching-based generation into representation learning, but relies on aggregation across encoder layers. The closest work, WavCube~\cite{WavCube}, follows a two-stage paradigm that adapts an encoder pre-trained for understanding into a joint semantic--acoustic model, with the original encoder as a semantic anchor. By contrast, the SPEAR-Gen representation encoder is trained from scratch, with its predictive and generative objectives jointly optimised.

\vspace{-0.4em}
\section{SPEAR-Gen}
\label{sec:spear_gen_methods}




\vspace{-0.4em}
\subsection{Task-Aligned Feature Aggregation for MTP}
\label{sec:tafa}

To address the information bias of MVQ tokens inherited from the single-layer representation, TAFA consolidates complementary linguistic and paralinguistic cues across layers into a task-aligned representation to construct MVQ targets.
Let $\{\mathbf{S}^{(\ell)}\}_{\ell=1}^{L}$ denote the layer-wise representations of a pre-trained SSL encoder. TAFA applies layer-specific normalisation, concatenates the resulting features along the feature dimension, and maps them into a shared space using a learnable linear projection $P_{\phi}$:
\vspace{-0.2em}
\begin{equation}
\mathbf{Z}
=
P_{\phi}\!\left(
\operatorname{Concat}\!\left(
\operatorname{LN}_{1}(\mathbf{S}^{(1)}),
\ldots,
\operatorname{LN}_{L}(\mathbf{S}^{(L)})
\right)
\right).
\vspace{-0.1em}
\end{equation}
TAFA is trained on speech--text pairs spanning linguistic and paralinguistic tasks, with the aggregated representation $\mathbf{Z}$ conditioning a text LLM. The alignment objective is $\mathcal{L}_{\mathrm{align}}=-\mathbb{E}_{j}\log p(y_j\mid y_{<j},\mathbf{Z})$. Only the TAFA parameters $P_\phi$ and the LLM LoRA parameters are updated, while the SSL encoder remains frozen to preserve its generalisation capability. Unlike task-specific weighted sums~\cite{superb}, which learn separate layer combinations for individual downstream tasks, TAFA learns a single fusion shared across tasks.


After task alignment, we quantise the frame-wise $\mathbf{Z} = \mathbf{z}_{1:t}$ into $N$ MVQ tokens, $\mathbf{q}_t=Q_{\mathrm{MVQ}}(\mathbf{z}_t) =(q_{t,1},\ldots,q_{t,N})$, for MTP pre-training. 
Let $\mathbf{X}$ and $\widetilde{\mathbf{X}}$ denote the clean and masked acoustic input, the SPEAR-Gen encoder $E_{\theta}$ maps $\widetilde{\mathbf{X}}$ to the output $\mathbf{H}=E_{\theta}(\widetilde{\mathbf{X}})=(\mathbf{h}_1,\ldots,\mathbf{h}_T)$, from which $N$ classification heads predict the corresponding MVQ tokens:
\vspace{-0.5em}
\begin{equation}
\mathcal{L}_{\mathrm{MTP}}
=
-\frac{1}{TN}
\sum_{t=1}^{T}
\sum_{n=1}^{N}
\log p_{\theta,n}
\left(q_{t,n}\mid\mathbf{h}_t\right).
\vspace{-0.4em}
\end{equation}
This objective directly shapes a single encoder output representation, without downstream layer aggregation.

\vspace{-0.55em}
\subsection{Two-Stage Coarse-to-Fine Generative Modelling}
\label{sec:generative}
\vspace{-0.2em}


Masked prediction does not explicitly require $\mathbf{H}$ to preserve acoustic detail useful for generation. We therefore combine an $\ell_1$ log-mel objective for dominant spectro-temporal structure with FM over its residual for finer acoustic variation.

\vspace{-0.5em}
\subsubsection{Log-Mel Reconstruction}
\vspace{-0.2em}

We use mean--variance normalised log-mel spectrograms
$\mathbf{M}\in\mathbb{R}^{T\times F}$ as reconstruction targets. Given encoder output $\mathbf{H}$, the mel decoder $D_{\mathrm{mel}}$ predicts $\widehat{\mathbf{M}}=D_{\mathrm{mel}}(\mathbf{H})$. The $\ell_1$ loss is computed between predicted and ground-truth mel:
\vspace{-0.4em}
\begin{equation}
    \mathcal{L}_{\mathrm{mel}}
    =
    \frac{1}{TF}
    \left\|
        \widehat{\mathbf{M}}-\mathbf{M}
    \right\|_{1}.
    \label{eq:mel_reconstruction}
\vspace{-0.25em}
\end{equation}
Since $D_{\mathrm{mel}}$ is conditioned solely on $\mathbf{H}$, $\mathcal{L}_{\mathrm{mel}}$ encourages the encoder output to retain dominant spectro-temporal structure.

\vspace{-0.3em}
\subsubsection{Residual Flow Matching}

An $\ell_1$ mel objective produces a single point estimate that tends towards the conditional median, potentially under-representing fine-grained acoustic variation. We therefore apply conditional FM to the residual of the mel prediction, providing a complementary learning signal that encourages $\mathbf{H}$ to retain acoustic cues not captured by the coarse prediction.

Specifically, we define the residual as $\mathbf{R}=\mathbf{M}-\operatorname{sg}(\widehat{\mathbf{M}})$. The stop-gradient separates the two stages by preventing the FM branch from updating $D_{\mathrm{mel}}$ through the residual target, while the encoder is still updated through conditioning on $\mathbf{H}$. Given $\boldsymbol{\epsilon}\sim\mathcal{N}(\mathbf{0},\mathbf{I})$, we define $\mathbf{R}_{\tau}=(1-\tau)\boldsymbol{\epsilon}+\tau\mathbf{R}$ and the target velocity $\mathbf{u}_{\tau}=\mathbf{R}-\boldsymbol{\epsilon}$. A conditional FM decoder $D_{\mathrm{FM}}$ takes $(\mathbf{R}_{\tau},\tau,\mathbf{H})$ as input and predicts $\mathbf{u}_{\tau}$ using
\vspace{-0.3em}
\begin{equation}
\mathcal{L}_{\mathrm{FM}}
=
\mathbb{E}_{\tau,\boldsymbol{\epsilon}}
\left[
\frac{1}{TF}
\left\|
D_{\mathrm{FM}}(\mathbf{R}_{\tau},\tau,\mathbf{H})
-\mathbf{u}_{\tau}
\right\|_{2}^{2}
\right]
\label{eq:residual_fm}
\vspace{-0.3em}
\end{equation}
We sample $\tau\sim\operatorname{Beta}(2,5)$, favouring noisier input to encourage $\mathbf{H}$ to model fine-grained acoustic detail.


\vspace{-0.3em}
\subsection{Unifying Predictive and Generative Objectives}
\label{sec:joint_objective}
\vspace{-0.2em}

SPEAR-Gen jointly optimises predictive and generative objectives, leading to the following pre-training loss:
\vspace{-0.3em}
\begin{equation}
    \mathcal{L}
    =
    \mathcal{L}_{\mathrm{MTP}}
    +
    \lambda
    \left(
        \mathcal{L}_{\mathrm{mel}}
        +
        \gamma\mathcal{L}_{\mathrm{FM}}
    \right),
    \label{eq:joint_objective}
\vspace{-0.3em}
\end{equation}
where $\lambda$ and $\gamma$ control the overall generative objective and residual FM, respectively. To reduce the mismatch between masked pre-training and clean-input inference, masking on $\mathbf{X}$ is bypassed with probability $p_{\mathrm{clean}}$.

\vspace{-0.2em}
\section{Experimental Setup}
\label{sec:exp_setup}

\vspace{-0.2em}
\subsection{Model and Pre-training Configuration}

SPEAR-Gen adopts Zipformer~\cite{zipformer} as its encoder architecture. The encoder receives 128-d log-mel filterbank features at 100~Hz and outputs representations at 50~Hz. The reconstruction target $\mathbf{M}$ is the same as the input feature $\mathbf{X}$. Two model configurations are considered: Base version contains 95M encoder parameters with a
512-d output, while Large contains 327M parameters with a 1024-d output. Both $D_{\mathrm{mel}}$ and $D_{\mathrm{FM}}$ use the same lightweight ConvNeXt~\cite{ConvNeXt} architecture, each containing approximately 15M parameters.

For task-aligned TAFA training, we use the frozen SPEAR-XLarge checkpoint as the SSL feature extractor and Qwen3-0.6B as the LLM decoder. The SPEAR-XLarge model has 13 layers with hidden size of 1280. TAFA concats the 13 hidden states and projects back to 1280-d features. Only TAFA and LoRA parameters in the LLM are updated. After task alignment, the 1280-d output of TAFA at 50~Hz frame rate is used to train an MVQ quantiser with eight 256-entry codebooks, which is then used to produce targets for MTP as introduced in Sec~\ref{sec:tafa}.
Following SPEAR, approximately 50\% input fbank frames are masked, while masking is bypassed with $p_{\mathrm{clean}}=0.1$. We set $\lambda=\gamma=0.5$. The global batch contains 4,800 seconds of speech. The Base and Large models are trained for 100k and 200k steps, respectively.

\vspace{-0.8em}
\subsection{Training Data}
\vspace{-0.2em}

Table~\ref{tab:training_data} summarises the datasets and task-label usage for TAFA alignment and SPEAR-Gen pre-training. Approximately 350 hours of labelled speech were used during TAFA alignment. These data are used exclusively for TAFA alignment, while SPEAR-Gen pre-training uses LibriSpeech for the Base model and LibriHeavy for the Large model without annotations. 
All TAFA alignment corpora are disjoint from the SUPERB evaluation datasets.
Thus, task supervision is introduced only through external TAFA alignment data, without exposing evaluation transcripts or speaker identities, allowing a fair evaluation of the TAFA contribution.

\begin{table}[t]
\centering
\vspace{-0.5em}
\caption{Datasets, task-label usage, and training objectives for
TAFA alignment and SPEAR-Gen pre-training.}
\label{tab:training_data}
\setlength{\tabcolsep}{2.5pt}
\begin{adjustbox}{max width=\columnwidth}
\begin{tabular}{@{}lllcc@{}}
\toprule
Stage & Dataset & Hours & ~Task labels~ & Objectives \\
\midrule
\multirow{2}{*}{\makecell[l]{TAFA\\alignment}}
& VoxPopuli~\cite{VoxPopuli} & 200
& \multirow{2}{*}{Yes} & ASR, SV\\
& MSP-Podcast~\cite{MSPPodcast} & 150 & & ER \\
\midrule
\multirow{2}{*}{\makecell[l]{SPEAR-Gen\\Pre-training}~~}
& LibriSpeech~\cite{librispeech} & 960
& \multirow{2}{*}{No}
& \multirow{2}{*}{MTP, Gen.} \\
& LibriHeavy~\cite{libriheavy} & 50k & \\
\bottomrule
\end{tabular}
\end{adjustbox}
\vspace{-1em}
\end{table}

\vspace{-0.6em}
\subsection{Evaluation Protocols}
\vspace{-0.1em}

We evaluate the final-layer representation along two complementary axes: speech understanding and speech resynthesis.

\noindent\textit{\textbf{Speech understanding.}}
We evaluate the frozen representations on SUPERB~\cite{superb}, using only the encoder's final layer. While the conventional SUPERB protocol learns a task-specific weighted sum over encoder layers, our protocol directly assesses whether diverse information is jointly accessible from a \textit{single task-independent representation}. The selected tasks cover linguistic information through 
automatic speech recognition (ASR), phoneme recognition (PR), keyword spotting (KWS) and intent classification (IC), as well as paralinguistic and speaker-related information through emotion recognition (ER), speaker identification (SID), speaker verification (SV), and speaker diarisation (SD). 

\noindent\textit{\textbf{Speech resynthesis.}}
To assess whether the same representation retains acoustic information useful for speech generation, we train a 14M HiFi-GAN~\cite{HiFiGAN} vocoder conditioned directly on final-layer representations. All systems use the original HiFi-GAN architecture and are trained with LibriSpeech. Only the input layer of the HiFi-GAN is adjusted to match the representation dimension. We evaluate on test-clean using PESQ-WB, ViSQOL~\cite{ViSQOL}, and UTMOS~\cite{UTMOS} for perceptual quality, STOI and Whisper large-v3 WER for intelligibility, and SIM for speaker preservation. SIM is computed as the cosine similarity between
ECAPA-TDNN~\cite{ECAPATDNN} embeddings of the original and resynthesised utterances.

\begin{table*}[t]
\centering
\caption{Final-layer results on SUPERB. Upper block: understanding-oriented models; lower block: unified understanding--generation models. PT data denotes representation pre-training data. Note: SPEAR-Gen uses external supervision only for TAFA target construction.
SPEAR$^\dagger$: reproduced version of SPEAR on LS-960. Best in bold and second-best underlined.}
\label{tab:understanding}
\footnotesize
\setlength{\tabcolsep}{2.5pt}
\renewcommand{\arraystretch}{1.05}
\begin{tabular*}{\textwidth}{@{\extracolsep{\fill}}lcc*{8}{c}@{}}
\toprule
& & \multirow{2}{*}{\makecell{\# Encoder\\Params}} & 
\multicolumn{4}{c}{SUPERB: linguistic} &
\multicolumn{4}{c}{SUPERB: paralinguistic / speaker} \\
\cmidrule(lr){4-7}\cmidrule(lr){8-11}
Method & PT data &  &
PR$\downarrow$ & ASR$\downarrow$ & KWS$\uparrow$ & IC$\uparrow$ &
ER$\uparrow$ & SID$\uparrow$ & SV$\downarrow$ & SD$\downarrow$ \\
\midrule
HuBERT Base~\cite{hubert}
& LS-960 & 95M
& 6.74 & 6.97 & 96.23 & 96.36
& 62.44 & 61.45 & 6.18 & 7.19 \\

WavLM Base~\cite{wavlm}
& LS-960 & 95M
& 5.59 & 6.42 & 96.36 & 96.57
& 61.35 & 47.04 & 8.01 & 6.41 \\

SPEAR$^\dagger$~\cite{spear}
& LS-960 & 95M
& \underline{4.09} & 3.90 & 96.85 & 98.37
& 65.78 & 55.47 & 6.80 & \underline{3.23} \\
\midrule
WavCube~\cite{WavCube}
& LS-960 & 350M
& 9.91 & 9.36 & 97.42 & 90.41
& 63.47 & 42.36 & {5.86} & 8.14 \\

SPEAR-Gen Base
& LS-960 & 95M
& 4.53 & \underline{3.71} & \underline{97.47} & \underline{98.73}
& \underline{69.48} & \underline{70.23} & \underline{5.08} & 3.26 \\

SPEAR-Gen Large
& LH-50k & 327M
& \textbf{3.25} & \textbf{3.41} & \textbf{97.70} & \textbf{99.21}
& \textbf{71.05} & \textbf{74.62} & \textbf{4.32} & \textbf{2.56} \\
\bottomrule
\end{tabular*}
\vspace{-1.7em}
\end{table*}

\section{Experimental Results}
\label{sec:exp_results}

\subsection{Understanding and Resynthesis Performance}
\vspace{-0.2em}



Table~\ref{tab:understanding} and ~\ref{tab:resynthesis_clean} present the SUPERB last-layer evaluation results and speech resynthesis results. 
Baseline results are obtained from official released checkpoints, except for SPEAR, which we reproduce on LS-960.
On SUPERB, SPEAR-Gen Base substantially improves SPEAR, its understanding-oriented baseline, on ASR, KWS, ER, SID, and SV, while showing only modest changes on PR, IC, and SD. 
These gains are accompanied by much stronger acoustic recoverability reflected by speech resynthesis: relative to SPEAR, PESQ increases from 1.20 to 2.94, ViSQOL from 1.70 to 4.12, and SIM from 0.467 to 0.892, while WER decreases from 2.61 to 2.22. 
Together, these results show that SPEAR-Gen substantially improves acoustic recoverability while maintaining strong understanding performance. Table 4 separates the contributions of TAFA targets and generative supervision.

Compared with WavCube, SPEAR-Gen Base achieves stronger results across all reported SUPERB tasks while providing comparable resynthesis quality with fewer encoder parameters. SPEAR-Gen Large further improves the SUPERB results and, on resynthesis, matches WavCube in PESQ and STOI while improving ViSQOL, WER, SIM and UTMOS. The combined results show that SPEAR-Gen provides a better balance between understanding performance and acoustic recoverability within a single representation.


\begin{table}[t]
\centering
\caption{Speech resynthesis results on LibriSpeech test-clean. 
}
\label{tab:resynthesis_clean}
\scriptsize
\setlength{\tabcolsep}{1.6pt}
\renewcommand{\arraystretch}{1.05}
\begin{tabular*}{\columnwidth}{@{\extracolsep{\fill}}lcccccc@{}}
\toprule
Method & PESQ$\uparrow$ & ViSQOL$\uparrow$ & UTMOS$\uparrow$ & STOI$\uparrow$ & WER$\downarrow$ & SIM$\uparrow$ \\
\midrule
WavLM Base~\cite{wavlm}          & 1.16 & 1.68 & 3.37 & 0.79 & 2.50 & 0.473\\
SPEAR$^\dagger$~\cite{spear} & 1.20 & 1.70 & 3.38 & 0.81 & 2.61 & 0.467 \\
WavCube~\cite{WavCube}             & 2.93 & 3.92 & \underline{3.72} & \underline{0.95} & 2.26 & \underline{0.900} \\
\midrule
SPEAR-Gen Base              & \underline{2.94} & \underline{4.12} & 3.69 & \underline{0.95} & \underline{2.22} & 0.892 \\
SPEAR-Gen Large             & \textbf{2.98} & \textbf{4.19} & \textbf{3.78} & \textbf{0.96} & \textbf{2.19} & \textbf{0.910} \\
\bottomrule
\end{tabular*}
\vspace{-1em}
\end{table}

\vspace{-0.7em}
\subsection{Ablation Studies}
\vspace{-0.2em}


We conduct ablations on the Base model trained on LibriSpeech. Table 4 separates the effects of MTP target construction and generative supervision. 
Holding generative supervision fixed, TAFA improves all reported metrics over single-layer targets, both with and without generative training. 
TAFA also consistently outperforms uniform layer averaging for MTP target construction, confirming the value of task-aligned fusion beyond aggregation alone. 
Holding TAFA targets fixed, adding generative supervision greatly enhances ViSQOL from 1.86 to 4.12 and SIM from 0.555 to 0.892, while improving ASR, ER, and SV. Conversely, generative-only training by removing MTP yields the highest ViSQOL and SIM but substantially degrades understanding, showing much weaker performance on SUPERB. These comparisons indicate that TAFA improves prediction targets, generative supervision strengthens acoustic recoverability, and MTP is important for preserving understanding.


\begin{table}[t]
\centering
\caption{Ablation of MTP target and generative modelling (Gen.). MTP target specifies the representation from which MVQ prediction targets are constructed. None denotes training without MTP. LayerAvg denotes uniform layer averaging. 
}
\label{tab:tafa_gen_ablation}
\scriptsize
\setlength{\tabcolsep}{1.2pt}
\renewcommand{\arraystretch}{1.05}
\begin{tabular*}{\columnwidth}
{@{\extracolsep{\fill}}lc*{6}{c}@{}}
\toprule
& & \multicolumn{3}{c}{SUPERB}
& \multicolumn{3}{c}{Resynthesis} \\
\cmidrule(lr){3-5}\cmidrule(lr){6-8}
MTP target & Gen.
& ASR$\downarrow$ & ER$\uparrow$ & SV$\downarrow$
& ViSQOL$\uparrow$ & WER$\downarrow$ & SIM$\uparrow$ \\
\midrule

Single-layer
& $\times$
& 3.90 & 65.78 & 6.80
& 1.70 & 2.61 & 0.467 \\

TAFA
& $\times$
& \underline{3.83} & \underline{68.32} & 6.15
& 1.86 & 2.54 & 0.555 \\

\midrule

None
& $\checkmark$
& 20.72 & 53.83 & 14.42
& \textbf{4.29} & 2.23 & \textbf{0.909} \\

Single-layer
& $\checkmark$
& 3.94 & 66.70 & 6.42
& 4.07 & \underline{2.24} & 0.866 \\

LayerAvg
& $\checkmark$
& 4.35 & 66.91 & \underline{6.02}
& 4.06 & \underline{2.24} & 0.873 \\

TAFA
& $\checkmark$
& \textbf{3.71} & \textbf{69.48} & \textbf{5.08}
& \underline{4.12} & \textbf{2.22} & \underline{0.892} \\
\bottomrule
\end{tabular*}
\vspace{-1.4em}
\end{table}

\begin{table}[t]
\centering
\caption{Ablation of the generative objectives. \#Params denotes the total number of parameters in decoders. 
}
\label{tab:generative_ablation}
\scriptsize
\setlength{\tabcolsep}{1.2pt}
\renewcommand{\arraystretch}{1.05}
\begin{tabular*}{\columnwidth}{@{\extracolsep{\fill}}lcccccccc@{}}
\toprule
& \multirow{2}{*}{\vspace{-0.5em}\makecell{\#Decoder\\Params}}& \multicolumn{3}{c}{SUPERB} & \multicolumn{4}{c}{Resynthesis} \\
\cmidrule(lr){3-5}\cmidrule(lr){6-9}
Gen. Obj. &  & ASR$\downarrow$ & ER$\uparrow$ & SV$\downarrow$ & PESQ$\uparrow$ & UTMOS$\uparrow$ & WER$\downarrow$ & SIM$\uparrow$ \\
\midrule
Mel $\ell_1$                       & 15M & 3.79 & 68.42 & \underline{5.41} & \underline{2.76} & 3.55 & \underline{2.25} & \underline{0.883} \\
Mel $\ell_1$                       & 30M & \textbf{3.70} & \underline{69.28} & 5.46 & 2.73 & \underline{3.61} & \textbf{2.22} & \underline{0.883} \\
Full-mel FM                        & 30M & 3.72 & 68.92 & 5.59 & 2.61 & 3.58 & 2.26 & 0.875 \\
\textbf{SPEAR-Gen}                 & 30M & \underline{3.71} & \textbf{69.48} & \textbf{5.08} & \textbf{2.94} & \textbf{3.69} & \textbf{2.22} & \textbf{0.892} \\
\bottomrule
\end{tabular*}
\vspace{-1.4em}
\end{table}

Table~\ref{tab:generative_ablation} examines the generative objectives. Doubling the mel decoder from 15M to 30M parameters yields mixed results, indicating that decoder capacity alone does not improve the learned representation.
Applying FM directly to the full mel target is also less effective for resynthesis. By contrast, the proposed coarse-to-fine objective provides the strongest overall balance between understanding and resynthesis, outperforming the capacity-matched mel decoder with only $\ell_1$ loss. These results support the effectiveness of residual FM beyond simply increasing decoder capacity.

\vspace{-0.7em}
\section{Conclusions}
\label{sec:conclusion}
\vspace{-0.2em}

We introduce SPEAR-Gen, a speech representation model that learns a single output representation for both speech understanding and generation. 
In SPEAR-Gen, task-aligned feature aggregation fuses complementary linguistic and paralinguistic cues into task-aligned prediction targets, while coarse-to-fine generative modelling encourages the same representation to preserve acoustic detail.
Experiments demonstrate that SPEAR-Gen effectively balances speech understanding and speech resynthesis within a single representation, and generative supervision even improves ASR, ER, and SV in the controlled ablation. 
Future work will investigate its suitability as a prediction target for text-conditioned speech generation.

\bibliographystyle{IEEEbib}
\bibliography{strings,refs}

\end{document}